\documentclass[final]{aipproc}

\layoutstyle{8x11double}

\begin{document}

\title 
      [A New, Simple Model for Black Hole QPOs]
      {A New, Simple Model for Black Hole High Frequency QPOs}


\author{L. Rezzolla}{
  address={SISSA, International School for Advanced Studies, Trieste, Italy},
  address={INFN, Sezione di Trieste, Trieste, Italy},
} 

\copyrightyear  {2004}

\begin{abstract}
Observations of X-ray emissions from binary systems have always been
considered important tools to test the validity of General Relativity in
strong-field regimes. The pairs and triplets of high frequency
quasi-periodic oscillations observed in binaries containing a black hole
candidate, in particular, have been proposed as a means to measure more
directly the black hole properties such as its mass and spin. Numerous
models have been suggested over the years to explain the QPOs and the
rich phenomenology accompanying them. Many of these models rest on a
number of assumptions and are at times in conflict with the most recent
observations. We here propose a new, simple model in which the QPOs
result from basic $p$-mode oscillations of a non-Keplerian disc of finite
size. We show that within this new model all of the key properties of the
QPOs: {\it a)} harmonic ratios of frequencies even as the frequencies
change; {\it b)} variations in the relative strength of the frequencies
with spectral energy distribution and with photon energy; {\it c)} small
and systematic changes in the frequencies, can all be explained simply
given a single reasonable assumption.\thanks{This work was in
collaboration with S'i. Yoshida, T. J. Maccarone and O. Zanotti}
\end{abstract}

\date{\today}

\maketitle

\bigskip	

	A number of models have been proposed to explain the high
frequency quasi-periodic oscillations (HFQPOs) seen in accreting black
hole systems \cite{03,04} and two of these seem particularly promising in
our view. The first model proposed, the discoseismic model \cite{05},
asserts that $g$-modes should become trapped in the potential well of a
Keplerian (i.e. geometrically thin) disc in a Kerr potential. The size of
the region where the modes are trapped depends on both the mass and the
spin of the accreting black hole.  Additional frequencies of oscillation
should be expected from pressure modes and corrugation modes. The
predictions of the model are well summarised by Kato \cite{07}. Given
pairs of high frequency QPOs and a proper identification of the
frequencies with the particular modes, one can measure both the black
hole mass and spin to relatively high accuracy
\cite{08}. The discoveries of three systems where the QPOs show a
harmonic structure with relatively strong peaks seen in integer ratios
\cite{09,04} $1:2$, $2:3$, or $1:2:3$ seem to cast some doubt upon this
model. However, the model remains viable for the intermediate frequency
QPOs in GRS 1915+105, seen at 67 Hz \cite{10} and 40 Hz \cite{09}. The
second model proposed, the parametric resonance model \cite{06}, asserts
that a harmonic relationship in the QPOs frequencies can be produced as a
result of resonances and was motivated by the small number integer ratios
of some of the QPOs seen from black holes candidates. In particular, the
parametric resonance model suggests that an initial turbulence spectrum
is amplified at a radius where the (radial) epicyclic frequency is in
resonance with the (azimuthal) orbital frequency, with the two
frequencies being in (small) integer ratios. These annuli tend to be
close to the black hole event horizon for the observed frequencies and,
hence, general relativistic effects can be important.  Given a
sufficiently accurate mass estimate for the black hole and a proper
identification of the ratio of the frequencies in resonance, again, a
black hole spin can be measured fairly accurately. It should be noted
that given the observed frequencies of 162 and 324 Hz in GRS 1915+105 it
is {\it not} possible to produce the 40 and 67 Hz QPOs with a
discoseismic model and the higher frequency QPOs with a resonance model
while retaining the same values for the black hole mass and spin
\cite{11}.

	A potential problem for both models is the observational evidence
for a frequency jitter in the HFQPOs of XTE~J1550-564, i.e. for small
variations of about 10\% in frequencies for about 15\% of the time. This
difficulty could be particularly severe for the parametric resonance
model, whose relevant frequencies confine the resonance to a narrow
region in radial coordinates (This situation is worsened for the 1:3
resonance for which the radial variation of the frequencies is even more
rapid \cite{04}.).

	We here propose an explanation for the high frequency QPOs in
black hole candidates that has distinct novel features and is based on a
{\it single} assumption, namely that a non-Keplerian disc orbits around a
rotating black hole. More specifically, we propose a model based on the
oscillation properties of a geometrically thick disc located in the
vicinity of a Kerr black hole. A more detailed discussion of this model
can be found in \cite{13}.

	In contrast to a Keplerian disc, a non-Keplerian disc (i.e. a
torus) is confined to a finite-size region determined uniquely by the
distribution of the specific angular momentum and by the pressure
gradients. In similarity with a star, however, restoring forces can be
used to classify different modes of oscillation in the disc.  A first
restoring force is the centrifugal force, which is responsible for
inertial oscillations of the orbital motion of the disc and hence for
epicyclic oscillations. A second restoring force is the gravitational
field in the direction vertical to the orbital plane and which will
produce a harmonic oscillation across the equatorial plane if a portion
of the disc is perturbed in the vertical direction. A third restoring
force is provided by pressure gradients and the oscillations produced in
this way are closely related to the sound waves propagating in a
compressible fluid. In a geometrically thick disc, the vertical and
horizontal oscillations are in general coupled and more than a single
restoring force can intervene for the same mode. Referring the reader to 
\cite{07} for a more detailed discussion, it is here sufficient to 
remind that $c$ modes are essentially controlled by the vertical
gravitational field, that $g$ modes are mainly regulated by centrifugal
and pressure-gradient forces, and that all of the restoring forces
discussed above play a role in the case of $p$ modes. It is also useful
to underline that because we are here interested in modes with a
prevalent horizontal motion and frequencies above the epicyclic one, we
are essentially selecting ``inertial-acoustic'' modes having centrifugal
and pressure gradients as only restoring forces. Hereafter we will refer
to these simply as $p$ modes.
	
	Following a recent investigation of the dynamics of perturbed
relativistic tori \cite{12}, we have analysed the {\it global oscillation
properties} of such systems and, more specifically, we have performed a
perturbative analysis of axisymmetric modes of oscillation of
relativistic tori in the Cowling approximation. The eigenvalue problem
that needs to be solved to investigate consistently the oscillation
properties of fluid tori is simplified considerably if the vertical
structure is accounted for by an integration in the vertical direction.
Doing so removes one spatial dimension from the problem and yields to
simple ordinary differential equations. While an approximation, this
simpler model captures most of the features relevant to the discussion of
$p$-modes in relativistic tori and is a first-step in a research field
which is in great part unexplored.

	We have solved the eigenvalue problem for a number of
vertically-integrated relativistic tori which have been perturbed through
the introduction of arbitrary pressure perturbations. The tori are
modelled as made of a perfect fluid obeying a polytropic equation of
state for an ultrarelativistic gas of electrons (other polytropic indices
have also been considered). The equilibrium models span a large parameter
space in which both the sizes and the physical properties (e.g. pressure
and density profiles, distributions of specific angular momentum,
polytropic indices) are varied considerably \cite{14}. The eigenfunctions
and eigenfrequencies found in this way correspond to global $p$-modes,
have been computed for the fundamental mode of oscillation as well as for
the first few overtones, and {\it all} have been found to be in a {\it
harmonic sequence} $2:3:4:\ldots$ to within a few percent.

\begin{figure}
  \includegraphics[height=.45\textheight]{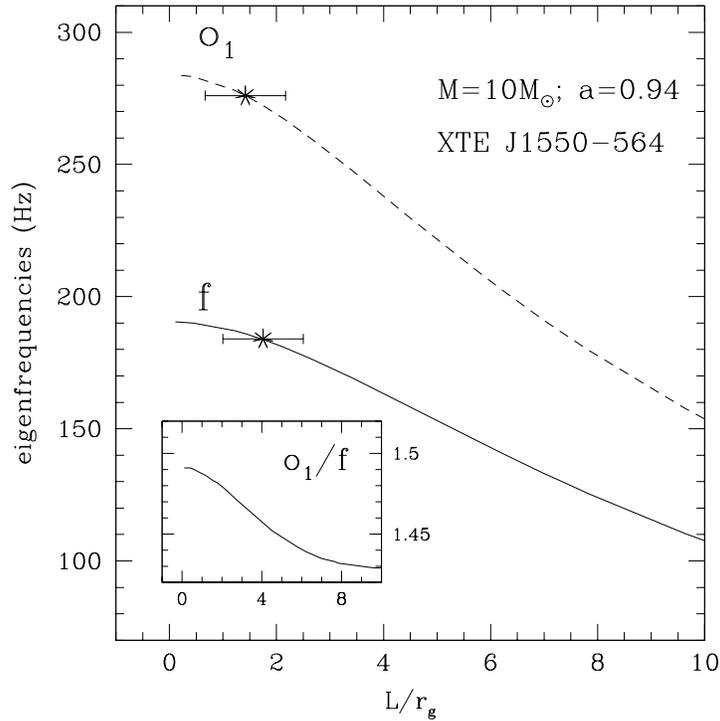}
  \caption{$p$-mode frequencies for a non-Keplerian disc. The horizontal
  axis shows the normalized radial dimensions of the torus $L/r_g$ when
  the black hole has a mass $M=10\ M_\odot$. The fundamental frequencies
  $f$ are indicated with a solid line, while the first overtones $o_1$
  with a dashed line. The filled dot indicates the epicyclic frequency at
  $r_{\rm max}$, while the two asterisks show the values of the HFQPOs
  observed in XTE~J1550-546. The error bars show the uncertainty in the
  determination of $M$.}
\label{fig1}
\end{figure}

        Figure~\ref{fig1} shows a typical example of the solution of the
eigenvalue problem for a black hole with mass $M=10\ M_{\odot}$ and its
comparison with the observations of XTE J1550-564 (other sources could
equally have been used). In particular, we have plotted the value of the
different eigenfrequencies found versus the radial extension $L$ of the
torus, expressed in units of the gravitational radius $r_g\equiv
GM/c^2$. The sequences have been calculated for a distribution of
specific angular momentum following a power-law \cite{13}, keeping the
position of maximum density in the torus at $r_{\rm max}=3.489$, and for
a black hole with dimensionless spin parameter $a \equiv J/M^2=0.94$ to
maximize the value for $L$. Indicated with a solid line are the
fundamental frequencies $f$, while the first overtones $o_1$ are shown
with a dashed line; each point on the two lines represents the numerical
solution of the eigenvalue problem. The asterisks represent the
frequencies of the HFQPOs detected in XTE~J1550-564 at 184 and 276 Hz,
respectively. Note that the two plotted eigenfrequencies are close to a
$2:3$ ratio over the full range of $L$ considered (this is shown in the
inset) and that while they depend also on other parameters in the problem
(e.g. the position of $r_{\rm max}$, the angular momentum distribution,
the polytropic index), these dependences are very weak so that the
frequencies depend effectively on $M$, $a$ and $L$. As a result, if $M$
and $a$ are known, the solution of the eigenvalue problem can be used to
determine the dimension of the oscillating region $L$ accurately.

	On the basis of these results, we suggest that the HFQPOs
observed in black hole candidate systems can be interpreted in terms of
$p$-mode oscillations of a small torus \cite{13}. Note that all the
properties discussed so far depend simply on the existence of
non-Keplerian disc orbiting in the vicinity of a black hole. Such a
configuration could be produced in a variety of ways and basically
whenever an intervening process [e.g. large viscosity, turbulence,
hydrodynamical and magnetohydrodynamical (MHD) instabilities] modifies
the Keplerian character of the flow.

	In what follows, we discuss how the predictions of such a model
can match the observed phenomenology and can be used to extract
astrophysical information.

\begin{description}

        \item {\it (i)} The harmonic relations between the HFQPO
        frequencies in black hole candidates are {\it naturally}
        explained within this model. In a sufficiently small torus, in
        fact, $p$ modes effectively behave as trapped sound waves with
        allowed wavelengths that are $\lambda = (2/2) L, (3/2)L, (4/2)L,
        \ldots$. The frequencies of these standing waves would be in an
        exact integer ratio only if the sound speed were constant. In
        practice this does not happen, but the eigenfrequencies found are
        in a sequence very close to $2:3:4$.

        \item{\it (ii)} Being global modes of oscillation, the same
        harmonicity is present at {\it all} radii within the torus. This
        removes the difficulty encountered in the resonance model and
        provides also a larger extent in radial coordinates where the
        emissivity can be modulated.

        \item {\it (iii)} Because the radial epicyclic frequency
        represents the upper limit for the disc eigenfrequencies, these
        scale like $1/M$. This is in agreement with the observations made
        of XTE~J1550-564 and GRO~J1655-40 as long as the spins are
        similar \cite{18}. On the other hand, a rather narrow range of
        black hole spins has been suggested as a possible explanation for
        the narrow range of radio-to-X-ray flux ratios in the Galactic
        X-ray binary systems \cite{20}.

        \item {\it (iv)} The frequency jitter can be naturally
        interpreted in terms of variations of the size of the oscillating
        cavity $L$. The frequencies may also drift over a large range
        with the harmonic structure preserved.

        \item {\it (v)} The observed variations in the relative strength
        of the peaks can be explained as a variation in the perturbations
        the torus is experiencing. (This has been reproduced with
        numerical simulations \cite{12}). Furthermore, while the low
        frequency overtones are energetically favoured and the
        corresponding eigenfunctions possess less nodes, {\it any number}
        of harmonics could in principle be observed.

        \item {\it (vi)} The evidence that an overtone can be stronger
        than the fundamental in the harder X-ray bands \cite{03}, can be
        explained simply given that the overtone is an oscillation
        preferentially of the innermost (and hottest) part of the
        accretion flow (see \cite{14} for the eigenfunctions).

\end{description}

A few remarks are worth making at this point. The first one is about the
existence of a non-Keplerian orbital motion: this is required {\it only
very close} to the black hole (the inner edge of the torus can be in
principle be located at the marginally bound orbit) and beyond this
region the fluid motion can be Keplerian. Stated differently, the HFQPOs
observed could be produced by the inner parts of an otherwise standard,
nearly-Keplerian, geometrically thin accretion disc, where a variety of
physical phenomena can introduce pressure-gradients (3D numerical
simulations in MHD seem to indicate the formation of these tori near the
black hole \cite{25}). The second remark is about the stability of these
tori to non-axisymmetric oscillations. It is well-known that a stationary
(i.e. non-accreting) perfect fluid torus flowing in circular orbits
around a black hole is subject to a dynamical instability triggered by
non-axisymmetric perturbations \cite{24}. It is less well-known, however,
that the instability can be suppressed if the flow is
non-stationary. Stated differently, a fluid torus around black could be
stable to non-axisymmetric perturbations if mass-accretion takes place
\cite{21,22,23}. Because the tori discussed here are assumed to be the
terminal part of standard accretion discs, we expect them to be stable to
non-axisymmetric perturbations as long as magnetic fields are
unimportant.

	Note that this model also offers a simple way of extracting
astrophysically relevant information from the observation of HFQPOs. We
recall that the fundamental $p$-mode frequency tends to the radial
epicyclic frequency $\kappa^2_{\rm r}$ at $r_{\rm max}$ in the limit of a
vanishing torus size, and that for a Kerr black hole this frequency is a
function of its mass and spin only. Exploiting this property, a first
estimate of the black hole spin and of the size of the oscillating torus
can be obtained once the lower frequency in the HFQPOs is measured
accurately. In this case, in fact, given a black hole candidate with
measured mass $M_*$, a {\it lower limit} on the black hole spin ${\bar
a}$ can be deduced as the value of $a$ at which the maximum epicyclic
frequency is equal to the lower observed HFQPO frequency
\cite{14}. Because for small tori $L \sim r_{\rm max}$, once
${\bar a}$ has been determined, the radial position ${\bar r}$ of the
maximum of the epicyclic frequency provides an {\it upper limit} on the
size of the torus, thus providing an estimate difficult to obtain through
direct observations. These two estimates should be considered first
approximations only and can be further refined through the eigenvalue
problem.

        A final comment will be reserved to the possibility of finding
support to this model from the observational data. There are now
extensive observations of these black hole HFQPOs and interesting
spectral and flux correlations are beginning to be found (see, for
instance, \cite{04}, and references therein). These observations could be
used to deduce the presence of a small torus in the terminal part of the
accretion discs. 

	Being based on a single assumption, our model for HFQPOs is
simply constructed, but, equally simply, it can be confuted. Two
observational constraints, if not met, will cast serious doubts about
this model. The first one requires the lower HFQPO frequency to be always
less than the maximum possible epicyclic frequency for a black hole of
mass $M_*$, i.e. lower HFQPO frequency$\;\leq{\rm max}[\kappa_{\rm r}(a =
1,M_*)]$. The second constraint, instead, requires that even if the
HFQPOs frequencies change as a result of the change in size of the torus,
they should nevertheless appear in a harmonic ratio to within
5-10\%. Both of these observational requirements are sufficiently
straightforward to assess and therefore provide a simple and effective
way of falsifying this model.

	Clearly, more work is needed to assess whether $p$-mode
oscillations in a small-size torus represent the correct interpretation
of the black hole QPO phenomenology. A first step in this direction will
be made when a convincing evidence is provided that $p$ modes can be
responsible for the modulated X-ray emission in black-hole spacetimes;
work is now in progress in this direction \cite{26}. We leave our final
remark on the complex phenomenology associated to HFQPOs in BHCs to a
well-known quote, whose view we feel to share: {\sl ``essentia non sunt
multiplicanda praeter necessitatem''} \cite{27}.

\begin{theacknowledgments}
It is a pleasure to thank Shin'ichirou Yoshida, Tom Maccarone and Olindo
Zanotti, my collaborators in this research.  Financial support has been
provided by the MIUR and by the EU Network Programme (European RTN
Contract HPRN-CT-2000-00137).
\end{theacknowledgments}



\begin{thebibliography}{50}
\expandafter\ifx\csname natexlab\endcsname\relax\def\natexlab#1{#1}\fi
\providecommand{\enquote}[1]{``#1''}
\expandafter\ifx\csname url\endcsname\relax
  \def\url#1{\texttt{#1}}\fi
\expandafter\ifx\csname urlprefix\endcsname\relax\def\urlprefix{URL }\fi

\bibitem[Strohmayer (2001a)]{03}~Strohmayer, T., 2001, {\it
ApJ}, {\bf 552}, L49

\bibitem[Remillard et al., (2002)]{04}~Remillard, R.A., Muno, M.P.,
McClintock, J.E. \& Orosz, J.A., 2002, {\it ApJ}, {\bf 580},1030

\bibitem[Nowak \& Wagoner (1991)]{05}~Nowak, M.A. \& Wagoner, R.V., 1991,
{\it ApJ}, {\bf 378}, 656

\bibitem[Kato (2001)]{07}~Kato, S., 2001, {\it Publ. Astron. Soc. Japan},
{\bf 53}, 1

\bibitem[Wagoner et al., (2001)]{08}~Wagoner, R.V., Silberlgleit, A.S. \&
Ortega-Rodriguez, M., 2001, {\it ApJ}, {\bf 559}, L25

\bibitem[Strohmayer (2001b)]{09}~Strohmayer, T., 2001b, {\it
ApJ}, {\bf 554}, L169

\bibitem[Morgan et al., (1997)]{10}Morgan, E.H., Remillard, R.A. \&
Greiner, J., 1997, {\it ApJ}, {\bf 482}, 993

\bibitem[Abramowicz \& Kluzniak (2001)]{06}~Abramowicz, M.A., \&
Kluzniak, W., 2001, {\it A\&A}, {\bf 374}, L19

\bibitem[Maccarone (2002)]{11}Maccarone, T.J., 2002, {\it
MNRAS}, {\bf 336}, 1371

\bibitem[Zanotti et al., (2003)]{13}Rezzolla L., Yoshida S'i., Maccarone
T. J., Zanotti L, 2003 {\it MNRAS}, 344, L37

\bibitem[Rezzolla et al., (2003)]{12}Zanotti O., Rezzolla L., Font J.A.,
2003 {\it MNRAS}, 341, 832

\bibitem[Zanotti et al., (2003)]{14}Rezzolla L., Yoshida S'i., Zanotti L, 
2003 {\it MNRAS}, 344, 978

\bibitem[Remillard et al., (2002)]{18}Remillard, R.A., Muno, M.,
McClintcock, J.E. \& Orosz, J., 2002, {\it ``New Views on Microquasars
''}, Eds. P. Duruchoux, Y. Fuchs \&
J. Rodriguez, p 49

\bibitem[Fender (2001)]{20}Fender, R.P., 2001, {\it
MNRAS}, {\bf 322}, 31

\bibitem[De Villiers \& Hawley (2003)]{25}{De Villiers, J.-P., Hawley,
J. F., 2003, {\it ApJ}, {\bf 592}, 1060}

\bibitem[Papaloizou \& Pringle (1984)]{24}{Papaloizou, J. C. B., Pringle,
J. E. 1984, {\it MNRAS} {\bf 208}, 721}

\bibitem[Blaes (1987)]{21}{Blaes, O., 1987, {\it
MNRAS}, {\bf 227}, 975}

\bibitem[Blaes \& Hawley (1988)]{22}{Blaes, O., Hawley, J. F., 1988, {\it
ApJ}, {\bf 326}, 277}

\bibitem[De Villiers \& Hawley (2002)]{23}{De Villiers, J.-P., Hawley,
J. F., 2002, {\it ApJ}, {\bf 577}, 866}

\bibitem[Karas \& Rezzolla (2004)]{26}{Karas, V., Rezzolla, L.  et
al. 2004, {\it in preparation}}

\bibitem[Occam (1330)]{27}{Occam, T., 1496 {\it Expositio aurea},
Bologna}

\end{thebibliography}
\end{document}